# EFFICIENT POWER THEFT DETECTION FOR RESIDENTIAL CONSUMERS USING MEAN SHIFT DATA MINING KNOWLEDGE DISCOVERY PROCESS

Blazakis Konstantinos[1] and Stavrakakis Georgios[2]

[1]School of Electrical and Computer Engineering, Technical University of Crete, Greece
*konst.blazakis@gmail.com*
[2]School of Electrical and Computer Engineering, Technical University of Crete, Greece
*gstavr@electronics.tuc.gr*

## ABSTRACT

*Energy theft constitutes an issue of great importance for electricity operators. The attempt to detect and reduce non-technical losses is a challenging task due to insufficient inspection methods. With the evolution of advanced metering infrastructure (AMI) in smart grids, a more complicated status quo in energy theft has emerged and many new technologies are being adopted to solve the problem. In order to identify illegal residential consumers, a computational method of analyzing and identifying electricity consumption patterns of consumers based on data mining techniques has been presented. Combining principal component analysis (PCA) with mean shift algorithm for different power theft scenarios, we can now cope with the power theft detection problem sufficiently. The overall research has shown encouraging results in residential consumers power theft detection that will help utilities to improve the reliability, security and operation of power network.*

## 1. INTRODUCTION

A substantial quantity of losses during the entire operation of the electrical network (generation, transmission and distribution), proves the involvement of non-technical losses, mainly in the distribution network, due to electricity theft by illegal consumers. Detection of illegal consumers is an extremely challenging problem nowadays, due to the large amount of money that is not imputable to the state and the electricity provider [1]-[7].

Technical losses in power systems are naturally occurring losses, which are caused by actions internal to the power system, and consist mainly of power dissipation in electrical system components such as transmission lines, power transformers and measurement systems [8], [9].

Non-technical Losses (NTLs) refer to losses that occur independently of technical losses in power systems. NTLs are caused by actions external to the power system, and also by the loads and conditions that technical loss computations fail to take into account. More specifically, NTLs are mainly related to power theft, and can also be viewed as undetected consumer loads that local utilities and electricity distribution companies don't know their existence.

NTLs are more difficult to measure because they are often non-countable by the system operators and thus have no recorded information [10]-[14]. There can be many reasons for power theft, such as high energy prices, unemployment, a weak financial situation of a consumer, tax purposes, weak accountability and enforcement of law, all of which reflect reasons to hide total energy consumption (moreover consumers who grow marijuana and produce drugs illegally or small-scale industries to hide a part or overall production) [15].

Techniques of power theft are also plentiful, such as using an external phase in front of meter terminals, tampering with meters so that meters record lower rates of consumption, inserting foreign materials and drilling holes into electromechanical energy meters, arranging false readings by manipulating meter readers as well as exposing the meter to mechanical shock. The most common and simplest way of pilfering electricity is tapping energy directly from an overhead distribution feeder [16], [17].

Monitoring of consumer load profiles for energy theft detection can be found in the literature [17]-[20]. The most common methods for fraud detection are Support Vector Machines ([21]-[23]), Artificial Neural Networks ([24], [39]), Bayesian Networks and Decision Trees [25], Extreme Learning Machines [26], Optimum-Path Forest [27], Fuzzy Clustering [28], Anomaly Detection [29] and Deep Learning which has recently achieved unprecedented performance in many areas of computer applications [30]-[32]. From the above technics, Support Vector Machines and Artificial Neural Networks are the leading technics due to good performance and easy adaption to different areas of research [17]-[19].

The main challenges and issues in NTL detection area are: handling imbalanced classes in the training data which also affects the evaluation metrics, describing features from the data which has a serious impact on the performance of a classifier, handling incorrect inspection results and recording results obtained through different methods comparable. This will allow researchers to deal with reliable, understandable and scalable results [17]-[19].

This paper presents a computational method that uses energy consumption measurement patterns to detect illegal residential consumers in a smart grid environment. In general, electronic meters (smart meters) collect real-time information from the consumers several times per day. Despite the numerous daily residential load profiles that appear ([33], [34]), due to the examination over a long time period an artificial dataset with near real-time energy consumption patterns has been developed in this work. Furthermore, the well-known clustering algorithm mean shift [35] along with a number of applications in the power theft detection field ([17]-[19], [36]-[38]) has been proposed and implemented, which in combination with PCA analysis successfully maps residential energy consumption patterns in terms of legal and illegal. The NTL methods based on artificial intelligence are typically applied to features computed from customer electricity consumption profiles and require the feature extraction from historical data for training process [46]. The computational effort for the feature extraction from historical data is not necessary when combining PCA data analysis with mean shift clustering algorithm. Given that the mean shift is an unsupervised clustering algorithm it can perform clustering to any amount of historical data available. Obviously, the more the historical electricity consumption data are, the more effective are the data clustering and the corresponding power theft detection results.

The main contributions of this paper are: a computational methodology for automated detection of illegal use of electricity in a local low voltage distribution network based on Principal Component Analysis (PCA) combined with the mean shift clustering, which is verified from the presented experimental results. Simulations with different numbers of power theft scenarios (Table 3), different percentages of partial power theft cases (scenario 2, Table 3), different percentages of power overload cases (power theft scenario 4, Table 3) and different consumption patterns for every consumer (Table 1). In this paper only the examination of power lines with non-technical losses is performed and not the whole power system (see Subsection 3.2), which is a computationally efficient fact. The insertion of smart meters via commercial power system analysis software is essential for the implementation of the proposed method in order to receive energy consumption measurements several times per day.

The rest of this paper is organized as follows. Section 2 provides an overview of how residential and commercial consumption patterns are constructed. Section 3 describes the power system model used in this research, the power theft scenarios implemented and what kind of data is received from a commercial power system analysis environment for further processing. Section 4 details the data preprocessing via data mining techniques (i.e. PCA, mean shift algorithm). In Section 5 experimental results are analyzed and finally a conclusion is presented in Section 6.

## 2. CONSTRUCTION OF RESIDENTIAL AND COMMERCIAL ELECTRICITY CONSUMPTION PATTERNS

Due to the lack of recorded statistics of the local utility from which we received data for an extended period of time, we were forced due to the needs of the present research to construct electricity consumption patterns for commercial and residential consumers.

For residential consumers, the length of the time period, during which people are presented within a dwelling, influences decisively the profile loads of residence, as the majority of electric devices requires human presence to be mobilized and supervised in order to maintain their working mode [40]-[41]. To determine the time periods where people are resident at home, the following frequent electricity consumption scenarios were used as shown in Table 1.

Table 1. Frequent electricity consumption scenarios of residential consumers

| Scenarios | Description |
|---|---|
| 1 | Absence from home 09:00 up to 13:00. Possibly inhabitants have part-time work in the morning. |
| 2 | Home absence from 09:00 up to 18:00. Possibly inhabitants have a full time work. |
| 3 | Home absence from 09:00 up to 16:00. |
| 4 | Full home presence. Possibly infant existence under people supervision or elderly people presence. |
| 5 | Home absence from 13:00 until 18:00. Possibly part time job during evening hours. |
| 6 | Full absence on weekdays and partial presence during weekends. Possibly cottage existence near to the permanent residence. |
| 7 | Almost complete home absence. Presence only some days of the year for holidays. Possibly existing cottage far away from the permanent residence. |

Except for the frequent electricity consumption scenarios mentioned, various combinations of these scenarios can be simulated. The number and the type of electrical device used on a daily basis at home is constantly changing. Nevertheless, the most intense loads are space heating, space cooling, water heating, refrigeration and lighting. The freezer is included in the base load because it is a permanent electricity consumer for the whole year, while the cooling load (heating, ventilation, air condition) is seasonal. The electrical loads that are included for residential consumption patterns are shown in Table 2.

Table 2. Load categories and electrical loads

| Load Categories | Electrical Loads |
|---|---|
| Personal hygiene | Water Heater, Hair Dryer |
| Preparation of food | Electrical oven, Microwave |
| Watching TV | Television |
| Heat, Cooling | Electrical Heating, Heat pump |
| Household chores | Vacuum cleaner, dishwasher, washing machine |
| Study | Computer |
| Base Load | Refrigerator, Freezer |
| Lighting | Electric lamps |

Based on our model, we examined the load profile over a six month period (April to September). The daily profiles are repeated every day, except for the weekends, by slightly changing the duration of activities and the magnitude of loads in order to produce different individual patterns. For the weekends, different profiles were constructed, as people tend to do outdoor activities. To include the influence of environmental conditions in the load profiles, we considered an increase in temperature in summer months. As a result, cooling loads were increased from the final load charts. Furthermore, it was considered that residents go on holidays once a year with a duration of 5 to 15 days.

Moreover, commercial patterns were constructed with the same methodology adjusting the electric loads and the duration of activities for every special business. Taking into account all the factors above, a database with load patterns for every electricity consumption scenario and different type of businesses was constructed. Load records account for every 15 mins, simulating the smart meters operation in a real environment. Fig. 1 and Fig. 2 show an example of a residential electricity consumption scenario 2 (Table 1) and respectively, restaurant power consumption per day. The increased consumption from 90 to 150 days (July-August) in Fig. 1 and Fig. 2 is due to the increased cooling load consumption as we previously noted. For residential consumers we consider a small reduction in cooling loads from the day 150 to the day 180 (i.e. during September), (see Fig. 1).

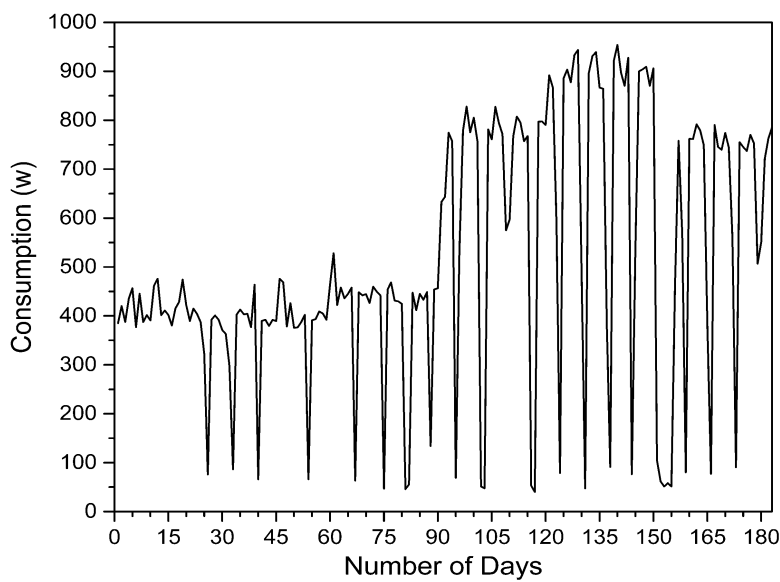

Figure 1. Scenario 2 power consumption from April to September

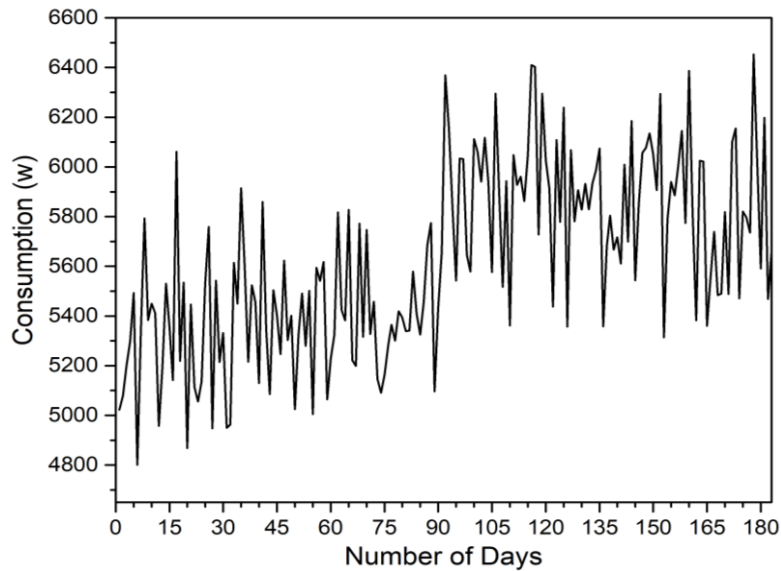

Figure 2.  Commercial power consumer (Restaurant)

# 3. MODEL DEVELOPMENT AND DATA EXTRACTION

## 3.1. Power system model

The power system model was developed by means of commercial power system analysis software using the educational version of 40 buses offered for free use. The commercial power system analysis software used is an interactive power system simulation package, designed to simulate power system steady state operation on a time frame ranging from several minutes to several days. The software contains a highly effective power flow analysis package, efficiently solving the power flow problem in power systems containing up to 250,000 buses.

Fig. 3 illustrates a part of the whole power system model, which is developed in a commercial power system analysis software environment for the needs of the present study. The power system model presented in this paper (i.e. Fig. 3) is a small part of the whole power system model, and consists of 3 commercial consumers (bold arrows) and 33 residential consumers (thin arrows). The whole power system model consists of 100 commercial and 1000 residential consumers. The lines are three phase balanced (one line equivalent) that can be modeled as a single line measuring between 250-300 meters in length. Each consumer has a fixed connection either single phase or three phase which is not possible to be changed by him.

In the proposed methodology the electricity consumption data at the consumer level are analyzed in order NTLs to be identified when unusual electricity consumption events are observed. Moreover, in the beginning of every line and for every consumer, we have installed the minimum number of sum meters and smart meters respectively, in order to examine the total power consumed over 6 months and for every 15 mins. At the moment, smart meters are able to measure the electricity consumption every minute, but, because of communicational system limitations, they send data once a day or once a week or even once a month. However, such measurements would be available in the near future. Moreover, in the real world, consumer smart meters do not work synchronously and because of communication delays, there is a slight time difference among the measured values recorded. This difference would be a source of error in the method proposed due to the fact that the method considers the measurements of all smart meters to be received at the same time instance. In any case, these delays occurring in the real world don't really affect the applicability of the proposed method. Along a line, we can observe that the number of consumers is increasing, as more and more consumers are connected in reality. The characteristics of the distribution lines are that of ACSR 50 R: 0.381 Ω/Km and X: 0.294 Ω/Km.

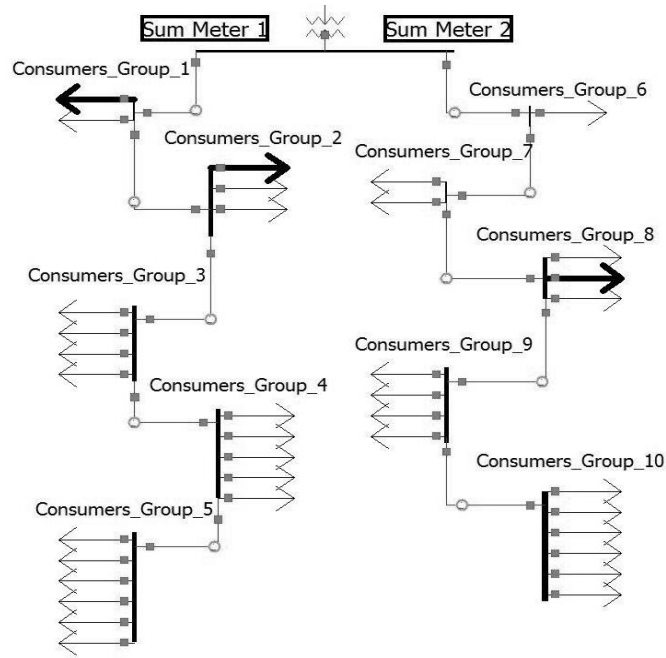

Figure 3. Power System Model in a commercial software environment

## 3.2. Power theft scenarios and local level power theft detection method

For the needs of the present work, we have constructed power theft scenarios as shown in Table 3.

Table 3. Power Theft Scenarios

| |
|---|
| 1. Consumers with smart meter fully stealing electricity due to power pass before the smart meter. |
| 2. Consumers with smart meter partially stealing electricity due to power pass before the smart meter. |
| 3. Consumers with no smart meter are connected illegally to the power grid. |
| 4. Consumers with abruptly increased consumption of electricity due to illegal activity or power delivery to unauthorized building. |

Having constructed a database for every frequent electricity consumption scenario (Table 1) and for different commercial consumers, as mentioned in Section 2, a random process ran for selecting a load pattern for every residential and commercial consumer in the grid. Consumption patterns were designed using Matlab software, and after they were introduced to the commercial power system analysis software used.

For our experiments, we apply first to our case study grid of Fig. 3, different combinations of the power theft scenarios (Table 3). In order to evaluate the proposed computational method, we apply power theft scenarios only in line 1 (Fig. 3 left line). More specifically we consider two residential consumers with power theft scenario 1, four residential consumers with power theft scenario 2, two residential consumers with power theft scenario 3 and three residential consumers with power theft scenario 4. Having introduced the smart meter readings for all consumers from our database to our case study grid, we ran our model for 6 months (April to September, i.e. from day zero to the day 180). We chose to receive results about the total active energy consumed from every line (sum meter reading), the total power losses for each line and the technical losses for each line respectively, as shown in Fig. 4. Total losses were estimated by the commercial power system analysis software, used for every line as the subtraction of consumers' consumption from sum meter readings. Subsequently, for every line, technical losses are subtracted from total losses so as to receive results for non-technical losses. We didn't take into consideration the reactive power because the percentage of reactive power for residential and commercial consumers was too low in contrast with industrial consumers. Fig. 4 shows the distribution line results.

More specifically, we can observe the sum meter readings in line 1 and line 2, total losses in line 1 and line 2, technical losses in line 1 and line 2, and non-technical losses in line 1 and line 2. As a result, non-technical losses in line 1 (Fig. 4 (d)) are far more remarkable in comparison with non-technical losses in line 2 (Fig. 4(h)). Non-technical losses in line 2 (Fig. 4 (h)) may be due to computational errors of the commercial power system analysis software used. Given that these errors are quite small, i.e. they are less than 10% of the technical losses in line 2, we didn't take them into consideration. Provided that significantly crucial technical losses are of priority for the electricity distribution companies, the result of the above observation leads to the detection of a power theft event in line one. The aim of the power theft detection method presented above is to permit electricity distribution companies to detect power theft events from the technical losses calculation of each line, which is easy to perform in real time by exploiting the advanced measuring features of the electronic smart meters.

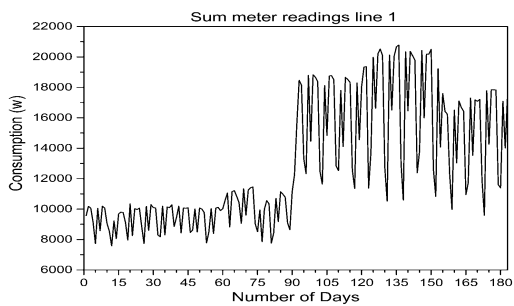

(a)

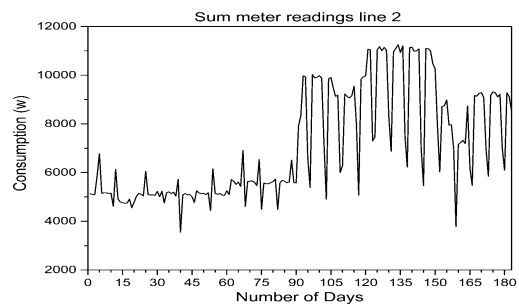

(e)

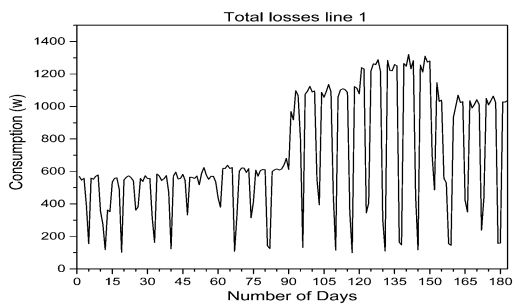

(b)

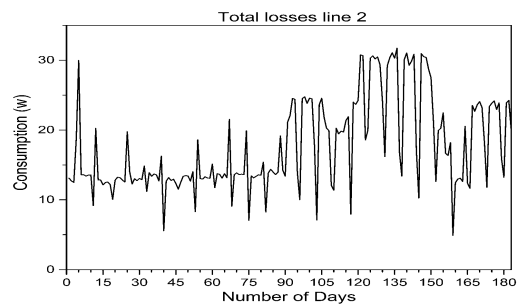

(f)

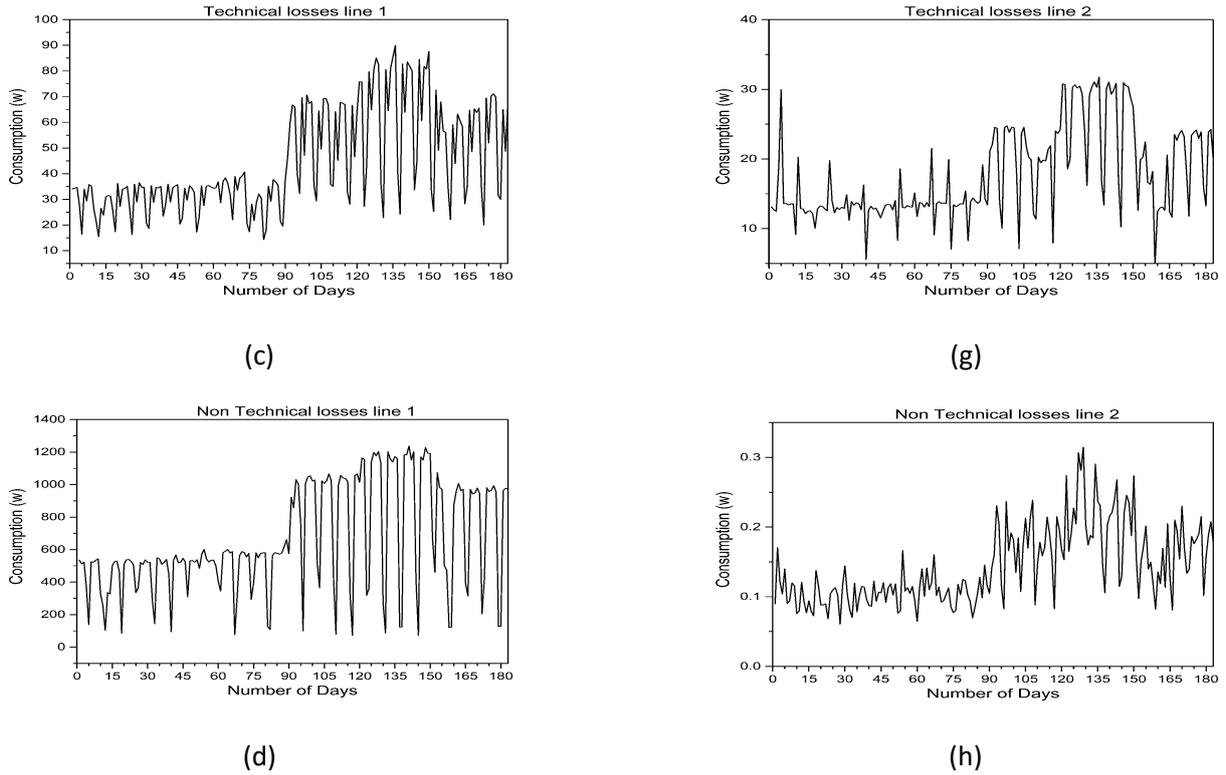

Figure 4. Distribution Line Results, (a) Sum meter readings in line 1, (b) Total losses in line 1, (c) Technical losses in line 1, (d) Non-technical losses in line 1, (e) Sum meter readings in line 2, (f) Total losses in line 2, ,(g) Technical losses in line 2, (h)Non-technical losses in line 2

## 4. Data Preprocessing

Having collected the appropriate results from the commercial power system analysis software used as mentioned in Section 3, the next step is to separate commercial and residential consumers from line 1 (Fig. 3 left line) a fact that can easily be done in reality, as the electricity provider knows the categories of the consumers-clients. From the residential consumers group, we remove those consumers with scenario 6, these with scenario 7 and those with zero consumption. For residential consumers, we keep only weekdays and we eliminate weekends due to the different habits of residents, especially during weekends. In this way, we enhance the clustering procedure to be more robust.

Subsequently, before applying principal component analysis (PCA) [42], [43] it is standard practice to first perform mean normalization at feature scaling, so that the features (power consumption) have zero mean and should have a comparable range of values due to the expression (1) below:

$$\mu_j = \frac{1}{m}\sum_{i=1}^{m} x_j^{(i)} \qquad (1)$$

where $X_j$ is the j$_{th}$ consumer data vector and $\mu_j$ the average electricity consumption of each consumer. Then we replace each $x_j^{(i)}$ with $(x_j^{(i)} - \mu_j)$.

Next step is to apply the technique principal component analysis (PCA) for dimensionality reduction while retaining most of the data's variance. PCA obtains eigenvalues and eigenvectors (principal components) which represent the characteristics and relationship of the data. Those characteristics with lower eigenvalues can be eliminated as they are not significant components.

In our survey, the dimensions of data are 17568 as it represents the number of data points "electricity consumption per 15 min and for 6 months". Subsequently we chose the first 10 principal components (data dimentions after the application of PCA) so as to keep the data variance to 95 % according to the expression (2):

$$\frac{\frac{1}{m}\sum_{i=1}^{m}\left\|x^{(i)}-x_{approx}^{(i)}\right\|^{2}}{\frac{1}{m}\sum_{i=1}^{m}\|x^{(i)}\|^{2}}\leq 0.05 \qquad (2)$$

In the expression (2) the numerator corresponds to the average square projection error with $x_{approx}^{(i)}$ denoting the projected data and the denominator, the total variation (m=17568, number of data points for a 6 month time period). After the principal component analysis, the implementation of the mean shift algorithm will take place [35].

Mean shift algorithm is a nonparametric clustering technique which does not require prior knowledge of the number of clusters, and does not constrain the shape of the clusters. Mean shift algorithm considers data points as a sample of a probability density function. If dense regions (or clusters) are present in the feature space, then they correspond to the mode (or local maxima) of the probability density function. For each data point, mean shift algorithm associates it with the nearby peak of the dataset's probability density function. For each data point, mean shift algorithm defines a window around it and computes the mean of the data point. Then it shifts the center of the window to the mean and repeats the algorithm till it converges. After each iteration, we can consider that the window shifts to a denser region of the dataset.

Mean shift assets are as follows: i) it does not presume spherical clusters, ii) it requires just one parameter (window size) to be tuned, iii) it finds a variable number of modes which are not given and it is robust to outliers and weak in non-constant regions, iv) it has no local minima, thus the clustering it defines is uniquely determined by the bandwidth, without the need to run the algorithm with different initializations, v) outliers, which can be very problematic for Gaussian mixtures and K-means, do not overly affect the kernel density estimates (KDE), other than creating singleton clusters.

Disadvantages of the mean shift algorithm are: i) the output depend on window size. A large window size might result in incorrect clustering and might merge distinct clusters, whereas a very small window size might result in too many clusters, ii) efficient implementation is used on short cuts in the search and it does not scale well directly with dimension of feature space when it is above ten, iii) the classic mean shift algorithm is time intensive. The time complexity of it is given by $O(Tn^2)$ where T is the number of iterations and n is the number of data points in the data set. For the implementation of the mean shift algorithm to the data of the present work, we define a kernel density estimator as in the expression (3):

$$f(x) = \frac{1}{nh^d}\sum_{i=1}^{n}K\left(\frac{x-x_i}{h}\right) \qquad (3)$$

with bandwidth $h > 0$, d the number of dimentions, $x_i$ the data points and kernel $K(\chi) = e^{-x^2/2\sigma^2}$ for the Gaussian kernel. Gausian kernels are easier to analyze and give rise to simpler formulas.

Mean shift algorithm can be considered to be based on the gradient ascent on the density contour. The generic formula for gradient ascent is:

$$x_1 = x_0 + \eta f'(x_0) \qquad (4)$$

$$\nabla f(x) = \frac{1}{nh^d} \sum_{i=1}^{n} K'\left(\frac{x - x_i}{h}\right) \qquad (5)$$

By applying the expression (5) to kernel density estimator and setting the expression (5) equal to zero, we get the expression (6).Expression (7) is the mean shift to a denser region [35]:

$$\vec{x} = \frac{\sum_{i=1}^{n} K'\left(\frac{x - x_i}{h}\right) \vec{x_i}}{\sum_{i=1}^{n} K'\left(\frac{x - x_i}{h}\right)} \qquad (6)$$

Assuming $g(x) = -K'(x)$ we have

$$m(x) = \frac{\sum_{i=1}^{n} g\left(\frac{x - x_i}{h}\right) \vec{x_i}}{\sum_{i=1}^{n} g\left(\frac{x - x_i}{h}\right)} - x \qquad (7)$$

The fundamental parameter in mean shift algorithm is the bandwidth h (window size), which determines the number of clusters. In the mean shift algorithm the number of clusters has no particular restrictions. Exploring a range of bandwidths, we ended up with a value that produces 5 clusters as the number of the most frequent residential consumers scenarios, i.e. the first five scenarios of Table 1. Scenario 6 and scenario 7 were removed during the data preprocessing. The cluster number chosen is 5 because the vast majoriy of consumers belong to the proposed first 5 classes. The systems don't know the residential type a priori. Each particular household is classified only by taking into account its consumption behaviour. The runtime of our experiments is less than a minute.

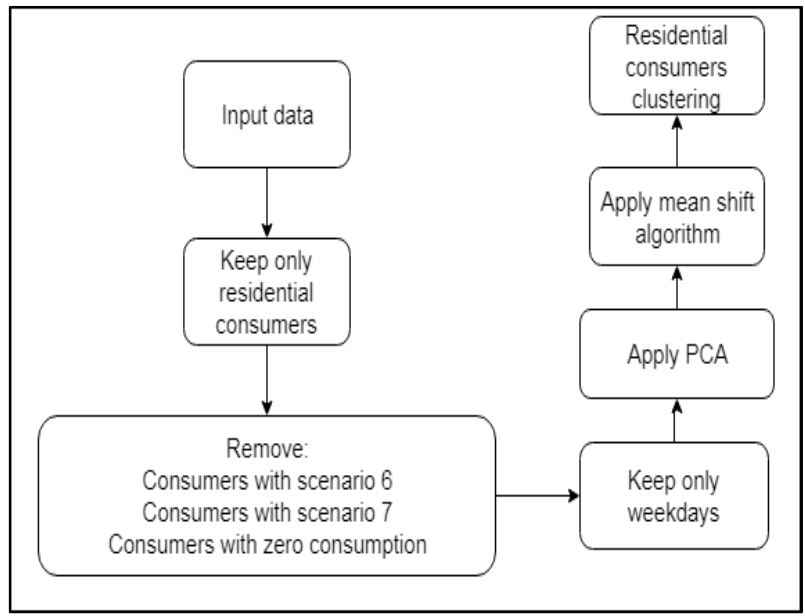

Figure 5.  Representation of the proposed method with a summarized flowchart consisted of 7 different steps

## 5. EXPERIMENTAL RESULTS

Simulating the whole system model using the commercial power system analysis software, which consists of 100 commercial and 1000 residential consumers, we randomly changed the number of power theft scenarios, the percentage of power theft scenario 2 (Table 3), the percentage of power overload consumption for power theft scenario 4 (Table 3) and the consumption patterns for every consumer. The power theft scenarios presented and examined here are the most frequently referred to and adopted in the bibliography [8].

Other scenarios that would be taken into account are power theft from a public organization, and power theft at specific times per day, but they are not considered in this work because they rarely occur.

As shown in Fig.6 (c) and Fig.6 (d), by simulating the whole system except for mean shift algorithm, but also with DBSCAN clustering algorithm [36], [44], for residential consumers, with partial power theft (power theft scenario 2) equal or higher to 65% of the usual consumption and for residential consumers with overload consumption equal or higher to 60% of the usual consumption (power theft scenario 4), we have high rates of success (hit rate metric) for power theft detection and power overload consumption detection for both algorithms. Hit rate (or sensitivity or recall or true positive rate (TPR)) measures the proportion of the number of True Positives divided by the number of True Positives and False Negatives and is defined as TP/(TP + FN), where TP is a consumer correctly identified and FN is a consumer incorrectly rejected [45]. The regular users of energy are the True Negative (TN) cases which are not necessary in the hit rate calculation.

For the zero consumption loads observed, (power theft scenario 1 or non-habitable dwelling) or for residential consumptions with scenario 6 and scenario 7 and for consumers with power theft scenario 3, the proposed detection method cannot be applied. In cases of zero consumption loads and for residential consumptions with scenario 6 and scenario 7, we need to verify if the house is inhabited or not by requesting the corresponding data from the tax services. In case the house is declared inhabited, then there is a strong indication for power theft incident.This step is necessary due to the fact that consumption scenarios 6 and 7 have very low consumption in comparison with the other scenarios.

In Fig. 6 (a) and Fig. 6 (b) we present, with respect to the first two most significant components of the PCA, the output of the mean shift algorithm and the output of DBSCAN algorithm respectively for a simulation with consumers with 20% power theft and 20% overload of the whole power system model. The 5 clusters which correspond to the 5 more frequent residential consumption scenarios, as well as the clusters with outliers which correpond to consumers with power theft scenario 2 and power theft scenario 4, are clearly shown in Fig. 6 (a) and Fig. 6 (b). For those outliers concerning consumers cases that do not belong to any class corresponding to the scenarios determined in Table 1 and Table 3, it must be suggested to the local utility to check any such outlier case individualy for potential power theft activity. In distribution networks of thousands of consumers, the clustering methodology presented in Section 4 permits us to classify the consumers according to the type (class) in which they belong to. Concerning residential consumers with power theft scenario 3, it can be said that it is impossible to be detected with the proposed method because no smart meters are connected to the power grid in that scenario. It is worth noting that other equally efficient metrics could be: AUC (area under curve), precision metric, F1 score, etc [18].

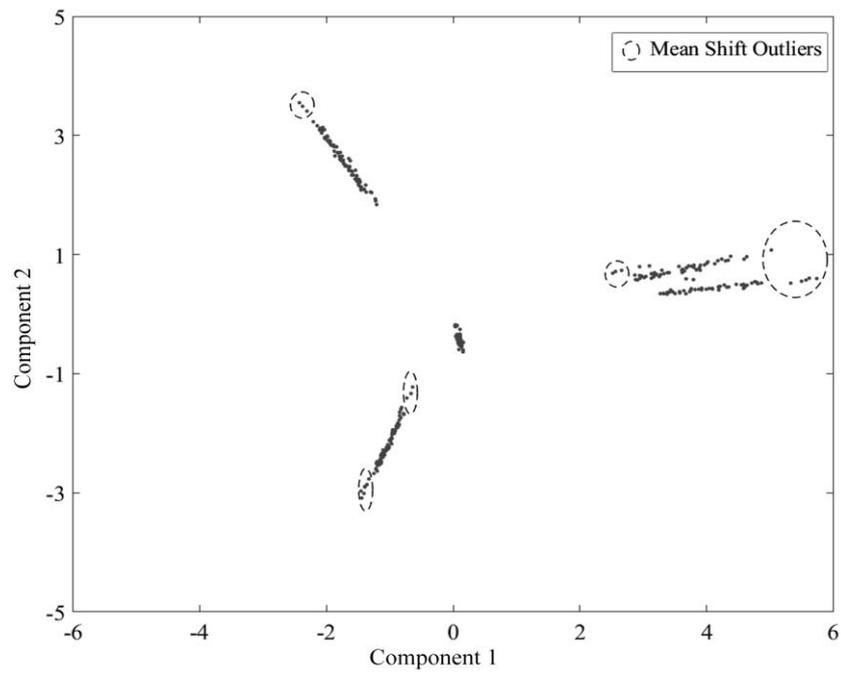

(a)

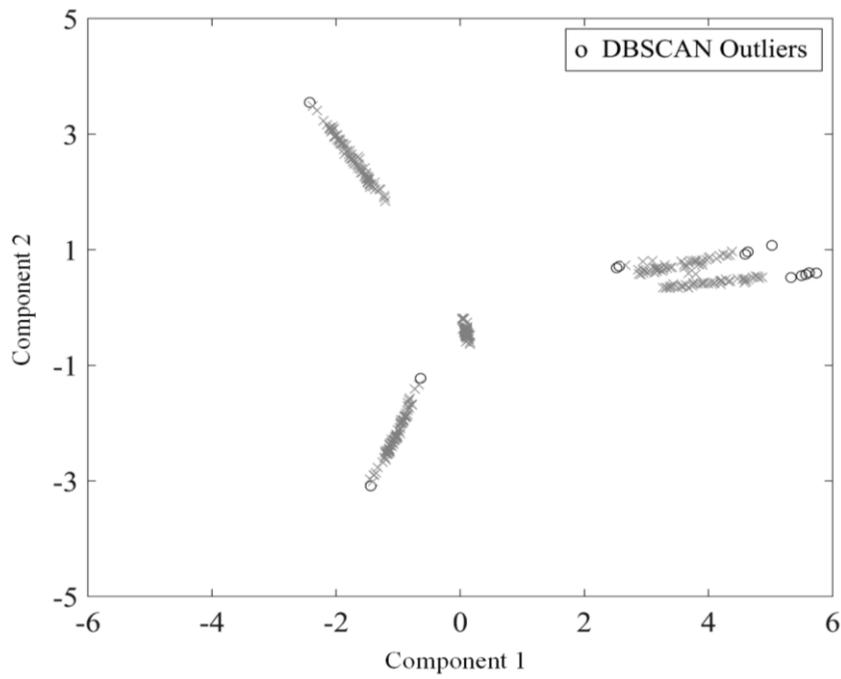

(b)

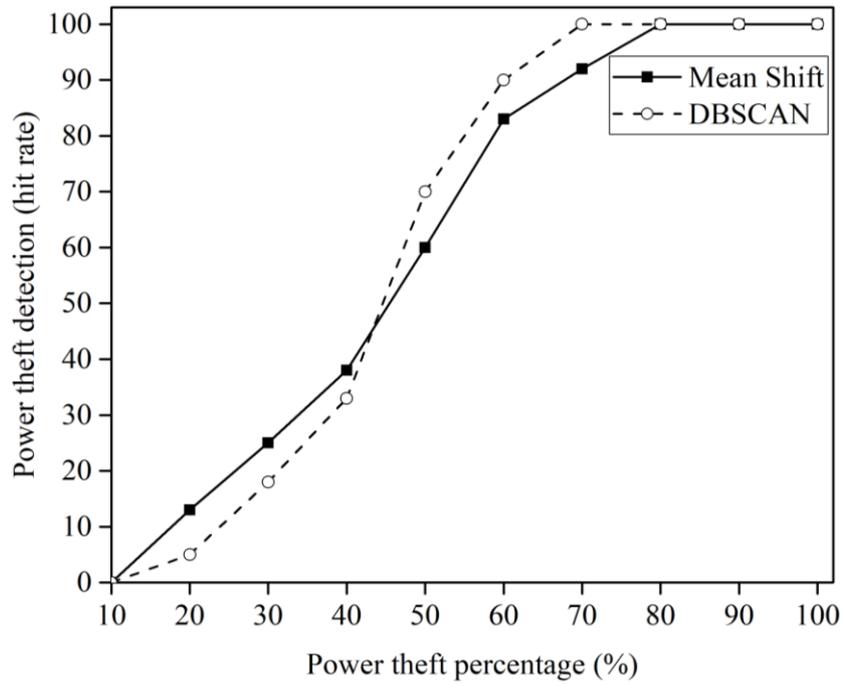

(c)

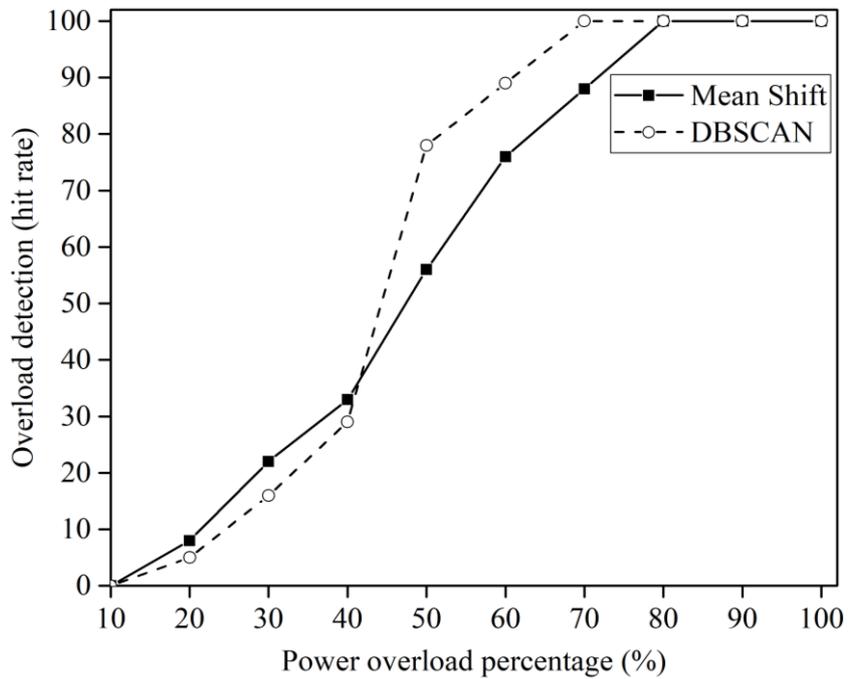

(d)

Figure 6.  Experimental results: a) Mean shift clustering results, b) DBSCAN clustering results, c) Mean shift-DBSCAN hit rate for power theft detection d) Mean shift-DBSCAN hit rate for abruptly increased power consumption detection

## 6. CONCLUSIONS

This paper initially presents the power theft phenomenon and demonstrates how much widespread it is among utilities worldwide. Energy-theft detection is a persistent and difficult problem to cope with in power grids management.

This paper reflects the fact that the development of advanced smart electronic metering infrastructure in smart grids allows for more manageable energy theft localization, thus advanced technologies of data mining can be adopted and developed in power theft successful detection. The presented methodology concerns only residential consumers, given that the commercial consumers present a very big variability in their electricity needs and consumption. Thus, the proposed methodology cannot be applied for commercial consumer power theft incidents.

Furthermore, the proposed methodology differs from other approaches in power theft case detection ([17]-[19]) in the examination and detection of various scenarios of partial power theft cases, with successful results as well as success in detecting power abrupt changes and overload consumptions, see Fig. 6. Moreover, the method does not need additional hardware installation, as it is based only on the data collected from electronic smart meters. In addition, knowing the consumption behaviour of a consumer (see Fig. 6. a), b)) we enhance applications in similar areas such as: power marketing strategy, public policy and social marketing.

A comparison among the accuracy of different solutions is difficult to establish due to the fact that studies deal with very different data, coming from different locations, representing different realities and presenting different data types [17]-[19]. By comparing mean shift algorithm with DBSCAN algorithm to the same data, it clearly shows in Fig. 6 (c) and Fig. 6 (d) that for power theft and power overload incidents with percentages less than 40% of the usual consumption, mean shift has a better performance compared with power theft and power overload percentages over 40% of the usual consumption where DBSCAN algorithm achieves higher success rates. Electricity theft detection should be considered as one of the most important aspects of efficient management in future distribution networks, as well.

In conclusion, the overall research has shown that the proposed methodology combines the advantages of the PCA dimensionality reduction of the big electricity consumption data with the advantages of the Mean-shift clustering algorithm, giving encouraging results for the successful partial power theft detection as well as for the illegal power overload detection in power distribution grids.

## ACKNOWLEDGEMENTS

This work was supported by H.F.R.I (Hellenic Foundation for Research and Innovation). This project has received funding from the Hellenic Foundation for Research and Innovation (HFRI) and the General Secretariat for Research and Technology (GSRT), under grant agreement No [81740].

**Authors**

**Blazakis Konstantinos:** Received his BSc degree in Applied Mathematical and Physical Sciences from N.T.U.A. (National Technical University of Athens) in 2010 and his MSc degree in Electrical and Computer Engineering from Technical University of Crete, Chania, Greece in 2015. Currently, he is a PhD candidate at Technical University of Crete. His areas of research include data mining, machine learning, smart grids, distributed electricity networks, renewable energy.

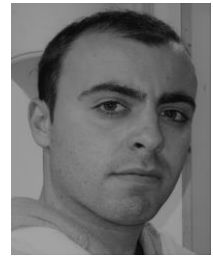

**Stavrakakis Georgios:** Received his first degree Diploma in Electrical Engineering from the N.T.U.A. (National Technical University of Athens), Athens, in 1980. His D.E.A. in Automatic Control and Systems Engineering was obtained from I.N.S.A., Toulouse, in 1981 and his Ph.D. in the same area was obtained from "Paul Sabatier-Toulouse III" University, Toulouse-France, in 1984. He has worked as a Research Fellow in the Robotics Laboratory of N.T.U.A. (1985-1988), and as a Visiting Scientist at the Institute for Systems Engineering and Informatics/Components Diagnostics & Reliability Sector of the Joint Research Center- EEC at Ispra, Italy (1989-1990). He was Vice President of the Hellenic Center for Renewable Energy Sources (CRES), Pikermi, Athens-Greece (2000-2002). He is currently a Full Professor at the Technical University of Crete, Chania, Greece.

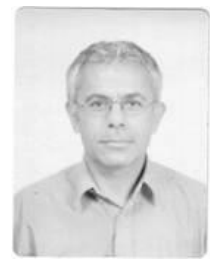